# Intrinsic Valley-Related Multiple Hall Effect in 2D Organometallic Lattice

Rui Peng, Zhonglin He, Ying Dai[*], Baibiao Huang, Yandong Ma[*]

School of Physics, State Key Laboratory of Crystal Materials, Shandong University, Shandanan Street 27, Jinan 250100, China

*Corresponding author: daiy60@sdu.edu.cn (Y.D.); yandong.ma@sdu.edu.cn (Y.M.)

Valley-related multiple Hall effect in 2D lattice is a fundamental transport phenomenon in the fields of condensed-matter physics and material science. So far, most proposals for its realization are limited to toy models or extrinsic effects. Here, based on tight-binding model and first-principles calculations, we report the discovery of intrinsic valley-related multiple Hall effect in 2D organometallic lattice of NbTa-benzene. Protected by the breaking of both time-reversal and inversion symmetry, NbTa-benzene exhibits large valley polarization spontaneously in both the conduction and valence bands, guaranteeing the anomalous valley Hall effect. Simultaneously, because of the exchange interaction and strong spin-orbit coupling, intrinsic band inversion occurs at one valley, which ensures the valley-polarized quantum anomalous Hall effect, thus presenting the extraordinary valley-related multiple Hall effect in nature. In addition, it can be transformed into the phase with ferrovalley or quantum anomalous Hall effect solely through strain engineering. These insights not only are useful for the fundamental research in valley-related physics, but also enable a wide range of novel device applications.

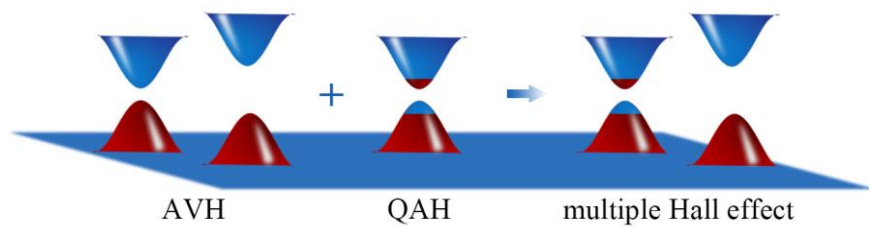

# I. Introduction

Valleytronics is a rapidly developing technique that aims to use the valley degree freedom as a medium for data storage and transfer [1-4]. Because of the large separation in momentum space, valley index is robust in terms of impurity and phonon scatterings [1-4]. A crucial ingredient for making exotic valleytronic device is valley polarization that correlates to time-reversal symmetry (***T***) breaking in two-dimensional (2D) materials with broken inversion symmetry (***P***) [5-20]. To realize the manipulation of valley index, several schemes have been proposed, including optical pumping [5-6], magnetic field/doping [7-8], proximity effect [9-10] and spontaneous valley polarization [11-20]. Among them, the later associated with intrinsic ***T***-breaking presents an intriguing case as it does not involve the shortcomings from external perturbations. Moreover, intriguing phenomena, such as the anomalous valley Hall (AVH) effect [11-20], related to spontaneous valley polarization has been observed. Nevertheless, so far, only a few 2D materials has been reported to exhibit spontaneous valley polarization [11-20], and most of them suffer from the annoying in-plane magnetization in nature [11-17].

Another striking phenomenon associated with intrinsic ***T***-breaking is the quantum anomalous Hall (QAH) effect [21-25]. The unique fingerprint of a QAH insulator is the existence of a chiral edge state in the insulating bulk gap, which is protected by ***T***-breaking and holds great potential for low-power-consumption device applications [21-25]. By coupling band topology to valley physics in QAH insulators, the valley-polarized QAH (VP-QAH) effect can be realized [26-38]. Assuming the AVH effect is preserved simultaneously, the long-sought valley-related multiple Hall effect is presented [26-27, 33-36]. Absolutely, compared with both AVH and VP-QAH effects, such multiple Hall effect is more intriguing as it raises the possibility to design dissipationless valleytronics as well as to explore novel fundamental physical phenomena [26-27, 33-36]. Despite the huge interest, the realization of VP-QAH effect usually accompanies with the deformation of AVH effect [28-32]. Therefore, the valley-related multiple Hall effect is rather scare, which is only proposed in a few systems based on theoretical models or extrinsic effect [26-27, 33-36]. Actually, up to now, intrinsic valley-related multiple Hall effect has not been reported yet.

In this work, combining tight-binding model with first-principles calculations, we identify the existence of intrinsic valley-related multiple Hall effect in 2D organometallic lattice of NbTa-benzene. NbTa-benzene is found to be a ferromagnetic (FM) semiconductor with band edges locating at the

±K points. Under the joint effect of ***T*-** and ***P*-**breaking, it harbors a sizeable valley polarization in both the conduction and valence bands spontaneously, which enables the observation of AVH effect. Moreover, arising from the exchange interaction and strong spin-orbit coupling (SOC), band inversion occurs at the +K valley naturally, while the band topology at the –K valley is trivial. This suggests the simultaneous existence of VP-QAH effect in NbTa-benzene, thus forming the exotic intrinsic valley-related multiple Hall effect. We further predict that under external strain, such multiple Hall effect can be derived into either AVH or QAH phase solely. The explored phenomena and mechanism greatly enrich the valley-polarized physics and highlight a promising platform to study valley-related multiple Hall effect.

**II. Methods**

First-principles calculations are performed based on density functional theory (DFT) [39] as implemented in Vienna ab initio simulation package (VASP) [40]. The Perdew-Burke-Ernzerhof (PBE) parametrization of generalized gradient approximation (GGA) [41] is used for describing exchange-correlation interaction. According to previous work [42], we adopt PBE+U method with a moderately effective $U_{eff} = (U - J) = 3.0$ eV for treating the strong correlation effect of d electrons of Nb and Ta atoms. The cutoff energy and electronic iteration convergence criterion are set to 450 eV and $10^{-5}$ eV, respectively. Structures are relaxed until the force on each atom is less than 0.01 eV/Å. A Monkhorst–Pack (MP) k-grid mess [43] of 5 × 5 × 1 is used to sample the 2D Brillouin zone. To avoid the interaction between adjacent layers, a vacuum space of 20 Å is added. For calculating Berry curvature and anomalous Hall conductivity, the maximally localized Wannier functions (MLWFs) implemented in WANNIER90 package [44] are employed. Edge states are obtained by WannierTools [45].

**III. Results and Discussion**

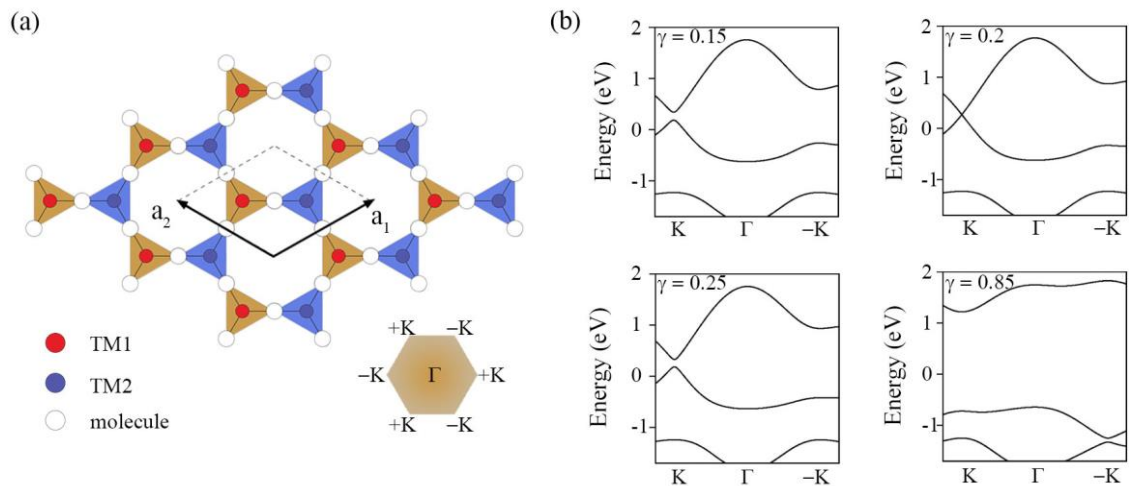

**Fig. 1** (a) Kagome lattice structure and its 2D Brillouin zone. (b) Tight-binding band structures of the Kagome lattice structure for $\gamma$ = 0.15, 0.20, 0.25 and 0.85 eV with $t$ = 1 eV and $\lambda$ = 0.35 eV.

Physically, both AVH and VP-QAH effects correlate to the energy extrema at the ±K point. The straightforward way to realize such energy extrema is to deform the Dirac point at the ±K point. For 2D systems harboring Dirac point, one typical example is the Kagome lattice, which is also known for its nontrivial properties. The unique fingerprint of Kagome lattice is the existence of two dispersive Dirac branches crossing around the Fermi level [25]. Upon breaking ***P***-symmetry, a band gap opens at the Dirac point, forming a pair of valleys at the ±K point. When further introducing ***T***-breaking in the Kagome lattice, the valley degeneracy would be lifted, leading to the possible realization of spontaneous valley polarization. Moreover, such degeneracy-lifting might yield different band orders at the +K and –K points. If these two features could coexist in such Kagome lattice, the valley-related multiple Hall effect can be achieved.

Following this principle, we construct a honeycomb-Kagome lattice assembled from the (quasi-)planer organic molecules (OM) and transition-metal (TM) atoms; see **Fig. 1(a)**. To satisfy ***T***-breaking, magnetic TM atoms are introduced. And to meet ***P***-breaking, the two TM atoms are chosen differently. We construct a three-band tight-binding model to investigate the band features around the Fermi level of such Kagome lattice. According to the lattice symmetry, the bands near the Dirac points are usually from the $p_z$ orbital of OM [25]. As the spin degeneracy would be lifted after considering magnetic exchange interaction, only one spin channel is considered in the model. The corresponding Hamiltonian of this Kagome lattice is given by

$$H(k_x, k_y) = \varepsilon_0 I + 2t \begin{pmatrix} 0 & \cos k_1 & \cos k_2 \\ \cos k_1 & 0 & \cos k_3 \\ \cos k_2 & \cos k_3 & 0 \end{pmatrix}$$

$$+ i2\lambda \begin{pmatrix} 0 & \cos k_1 & -\cos k_2 \\ -\cos k_1 & 0 & \cos k_3 \\ \cos k_2 & -\cos k_3 & 0 \end{pmatrix} + i2\gamma \begin{pmatrix} 0 & \sin k_1 & \sin k_2 \\ -\sin k_1 & 0 & \sin k_3 \\ -\sin k_2 & -\sin k_3 & 0 \end{pmatrix},$$

where $k_1 = \frac{\vec{a}_1}{2} \cdot \vec{k}$, $k_2 = \frac{\vec{a}_2}{2} \cdot \vec{k}$, $k_3 = \frac{(\vec{a}_2 - \vec{a}_1)}{2} \cdot \vec{k}$ and $\vec{k} = (k_x, k_y)$. $\varepsilon_0$ is the on-site energy, $t$ is the effective nearest-neighbor hopping parameter, $\lambda$ is the strength of intrinsic SOC and $\gamma$ represents Dresselhaus spin splitting. By diagonalizing the Hamiltonian, the band gap at ±K point can be expressed as

$$E_\tau = \frac{1}{2}(-3 + 3\tau\gamma + \sqrt{3}\lambda),$$

Here, $\tau = \pm 1$ represents valley index at ±K point. Interestingly, the valley polarizations are calculated to be $3\gamma/2$ in both the conduction and valence bands, implying the AVH effect in such Kagome lattice.

Moreover, we find that different values of $\gamma$ are associated with diverse phenomenon for this Kagome lattice. To illustrate it more clearly, we set $t$ = 1 eV and $\lambda$ = 0.35 eV, and plot the band structures with $\gamma$ = 0.15, 0.20, 0.25 and 0.85 eV in **Fig. 1(b)**. It can be seen that when $\gamma$ is less than 0.2 eV, the Kagome lattice only exhibit a spontaneous valley polarization, with a trivial band order at ±K valley. Upon increasing $\gamma$ to 0.2 eV, the band gap at +K valley is closed, while the sizeable band gap at –K valley is retained, forming a half-valley-metal (HVM) phase. When further increasing $\gamma$, a band gap reopens at +K valley, indicating a band inversion. While the band order remains trivial at –K valley, the VP-QAH is realized. Additionally, the valley features are maintained well after the band inversion for both the conduction and valence band, which suggests the preservation of the AVH effect. In this case, the system could exhibit the valley-related multiple Hall effect. When increasing $\gamma$ to 0.85 eV, the valence band maximum (VBM) moves from the +K to Γ point, deforming the AVH effect based on valence band. To confirm band topology of these case, we also calculate the Chern numbers $c$ at $\gamma$ = 0.15, 0.20, 0.25 and 0.85 eV, which are found to be $c$(0.15) = $c$(0.20) = 0 and $c$(0.25) = $c$(0.85) = 1. Therefore, by modulating the Dresselhaus spin splitting, the intrinsic valley-related multiple Hall effect indeed can be realized in the Kagome lattice.

One of the candidate Kagome systems to study the intrinsic valley-related multiple Hall effect is NbTa-benzene. **Fig. 2(a)** displays the crystal structure of NbTa-benzene. Each unit cell of NbTa-benzene consists of two TM atoms (one Nb and one Ta) and three benzene molecules. While the former forms a honeycomb lattice, the latter generates a Kagome lattice. The organic molecules are known to form strong bonds with TM atoms. Note that similar 2D metal-organic honeycomb-Kagome lattices have been successfully synthesized in experiment [46-47]. The lattice constant of NbTa-benzene is optimized to be 12.09 Å. Due to the strong steric repulsion, Nb and Ta atoms, respectively, move up and down along the out-of-plane direction, and the benzene rings rotate around the Nb-Ta axis, yielding a buckled structure with a vertical height of 1.32 Å. The space group of NbTa-benzene is $C_3^1$, which indicates its intrinsic *P*-breaking.

The electronic configuration of isolated Nb (Ta) atom is $4d^45s^1$ ($5d^36s^2$). By forming the Kagome lattice, three electrons are transferred from each TM atom to the neighboring benzene rings, producing the electronic configurations of $4d^25s^0$ and $5d^26s^0$, respectively, for Nb and Ta atoms. This would give rise to a magnetic moment of 2 $\mu_B$ for each Nb/Ta atom. As expected, our calculations show that the magnetic moment on each TM atom is 2 $\mu_B$. To determine the magnetic coupling between these magnetic moments, we consider the FM and antiferromagnetic (AFM) states. The FM state is found to be 0.19 eV/unit_cell lower than the AFM state, suggesting that it is a FM system. Therefore, *T*-breaking is also realized in the Kagome lattice. The magnetocrystalline anisotropy energy (MAE) of NbTa-benzene is calculated to be 0.18 eV/unit_cell, indicating its magnetization orientation is along the out-of-plane direction, which is an important premise for valleytronic systems [17-20].

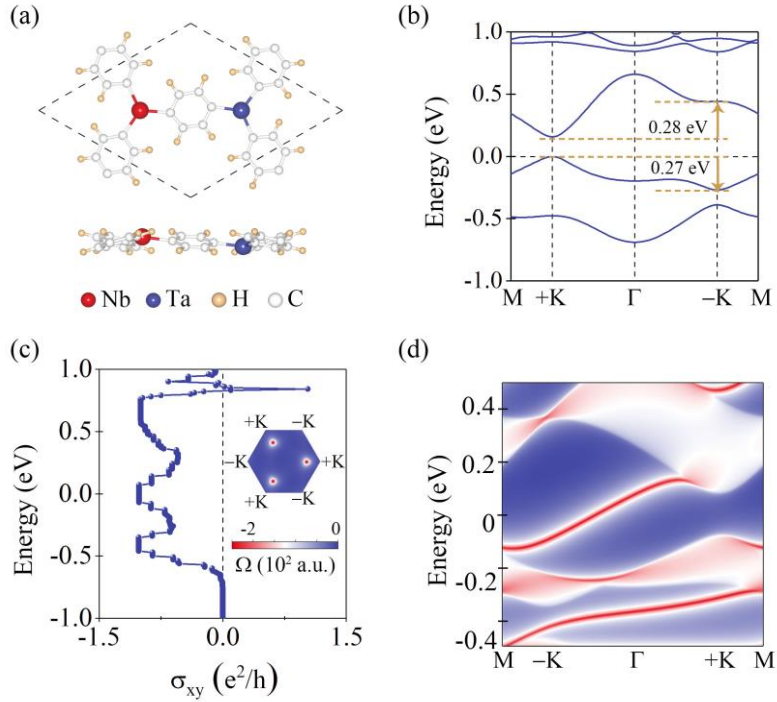

**Fig. 2** (a) Crystal structure of NbTa-benzene. (b) Band structure of NbTa-benzene with SOC. (c) Anomalous Hall conductivity of NbTa-benzene. Inset in (c) is the contour map of Berry curvature of NbTa-benzene over the entire 2D Brillouin zone. (d) Edge state of NbTa-benzene. The Fermi level is set to 0 eV.

**Fig. S1(a)** shows the spin-polarized band structure of NbTa-benzene without considering SOC. It can be seen that the spin-up and spin-down bands are largely separated in energy. For the low-energy states, they are mainly from the spin-down channels. NbTa-benzene exhibits a direct band gap of 0.25 eV, suggesting that it is a FM semiconductor. More importantly, as shown in **Fig. S1(a)**, its band edges are located at the ±K points. Considering its intrinsic *P*-breaking, this leads to a pair of energetically degenerate valleys in both the conduction and valence bands. By further considering SOC, the degeneracy between the +K and –K valleys is lifted [**Fig. S1(b)**], which is correlated to the intrinsic *T*-breaking. As a result, without employing complicated external methods, the valley polarization is realized spontaneously in NbTa-benzene. As shown in **Fig. 2(b)**, the valley polarization between +K and –K valleys is 0.28 and 0.27 eV in the conduction and valence bands, respectively. These values are much larger than those reported in previous works such as $LaBr_2$ (33 meV) [18] and $Nb_3I_8$ (107 meV) [16].

It is worth noting that the spontaneous valley polarizations in most of the existing systems are usually sizeable in either the conduction or valence band, but not both [11-20]. This is different from the case of NbTa-benzene, wherein the spontaneous valley polarization is large in both the conduction and valence bands. In this case, the valley index in both the conduction and valence bands can be utilized, highly desirable for valleytronic applications. To get physical insight into this phenomena, we investigate the orbital contribution to the band edges. As shown in **Fig. S2(a,b)**, the valleys in both the conduction and valence bands are dominated by TM-$d_{xz}$/$d_{yz}$ and C-$p_z$ orbitals. For C-$p_z$ orbital, because of its out-of-plane orientation character, it has almost no impact on the valley spin splitting and thus valley polarization. While for the TM-$d_{xz}$/$d_{yz}$ orbital, the strong SOC strength within the TM atoms results in the substantial valley spin splitting (see **Note** in Supplementary Information), leading to the giant valley polarization. As the conduction and valence bands of NbTa-benzene share similar orbital contributions, the sizeable valley polarization is observed in both the conduction and valence bands.

To explore the valley-contrasting physics in NbTa-benzene, we calculate its Berry curvature on the basis of the Kudo formula [48]:

$$\Omega(k) = \sum_n f_n \Omega_n(k)$$

$$\Omega_n(k) = -\sum_{n\neq n'} \frac{2Im\langle\psi_{nk}|v_x|\psi_{n'k}\rangle\langle\psi_{n'k}|v_y|\psi_{nk}\rangle}{(E_n - E_{n'})^2},$$

Here, $f_n$ is the Fermi-Dirac distribution function, $\psi_{nk}$ is the Bloch wave function with eigenvalue $E_n$, and $v_{x/y}$ is the velocity operator along $x/y$ direction. The contour map of Berry curvature of NbTa-benzene is shown in the insert in **Fig. 2(c)**. It can be seen that the Berry curvature of NbTa-benzene at –K valley is negligible, while it has a large negative value at +K valley. When shifting Fermi level between the +K and –K valleys in the conduction (valence) band, because of the Berry curvature, the spin-down (up) electrons (holes) in NbTa-benzene from the +K valley will acquire a transverse velocity and be accumulated at one edge of the sample under an in-plane electric field. Upon reversing magnetization orientation of NbTa-benzene, the –K valleys will be closer to the Fermi level; the Berry curvature at the +K will be negligible, and –K valleys will have a large positive value. In this case, the spin-up (down) electrons (holes) from the –K valley are accumulated at the opposite edge of the sample under an in-plane electric field. These result in the AVH effect in NbTa-benzene. Moreover, the measurable charge Hall current formed by the accumulated electrons (holes) from one valley can be detected as a voltage, which is applicable to realize valley-based data storage and transfer. To confirm this point, we calculate the anomalous Hall conductivity σ using the following equation: [49]

$$\sigma = \frac{e^2}{2\pi h}\sum_n \int d^2k \Omega_n.$$

As shown in **Fig. 2(c)**, when Fermi level is shifted between +K and –K valleys in the conduction (valence) band, a fully valley-polarized Hall conductivity will be generated.

From above, we confirm the AVH effect in NbTa-benzene. Except for this intriguing phenomenon, NbTa-benzene also harbors the nontrivial band topology. As shown in **Fig. S2(a)**, when excluding SOC, both +K and –K valleys in the conduction band are mainly contributed by the $d_{xz}/d_{yz}$ orbitals of Ta atoms (Ta-$d_{xz}/d_{yz}$) and $p_z$ orbitals of C atoms coordinated to Nb atoms ($C_{Nb}$-$p_z$), and those in the valence band are dominated by the $d_{xz}/d_{yz}$ orbitals of Nb atoms (Nb-$d_{xz}/d_{yz}$) and $p_z$ orbitals of C atoms coordinated to Ta atoms ($C_{Ta}$-$p_z$). Upon including SOC, as shown in **Fig. S2(b)**, the band order between Nb-$d_{xz}$/$d_{yz}$ & $C_{Ta}$-$p_z$ and Ta-$d_{xz}$/$d_{yz}$ & $C_{Nb}$-$p_z$ at +K valley is reversed. While for the band order at the –K valley, it remains trivial. Such valley-contrasting band inversion indicates the existence of VP-QAH effect in NbTa-benzene. To confirm this nontrivial band topology, the topological invariants of C is first calculated using [50]:

$$C = \frac{1}{2\pi}\sum_n \int d^2k \Omega_n$$

By integrating the Berry curvatures throughout the whole first BZ around each individual valley, the Chern number of NbTa-benzene is found to be C = -1, establishing the VP-QAH effect in NbTa-benzene. Furthermore, the clear plateau of σ = -$e^2$/h around the Fermi level displayed in **Fig. 2(c)** is also in consistent with the Chern number C = -1. Additionally, it can be seen from **Fig. 2(c)** that there are another two plateaus, which are related to the extra valley-contrasting band inversions above and below the Fermi level, respectively. However, as those two valley-contrasting band inversions are located far away from the Fermi level, they are useless and thus will not be addressed here. The corresponding edge states of NbTa-benzene are presented in **Fig. 2(d)**. Obviously, these is a gapless chiral edge state connecting the bulk conduction and valence bands at the +K valley, while the –K valley remains insulating, again confirming the VP-QAH effect in NbTa-benzene.

It is worth emphasizing that the realization of VP-QAH effect usually accompanies with the deformation of AVH effect [28-32]. As a result, the valley-related multiple Hall effect is rather scare. Remarkably, as shown in **Fig. 2(b)**, the appearance of VP-QAH does not deform the characterized valley features in NbTa-benzene, thus preserving the AVH effect. The simultaneous existence of AVH and VP-QAH phases gives rise to the long-sought intrinsic valley-related multiple Hall effect in NbTa-benzene, which agrees with the tight-binding model analysis. Given its large valley polarizations, such intrinsic multiple Hall effect in NbTa-benzene is highly desirable for designing dissipationless valleytronics as well as to explore novel fundamental physical phenomena [26-27, 33-36].

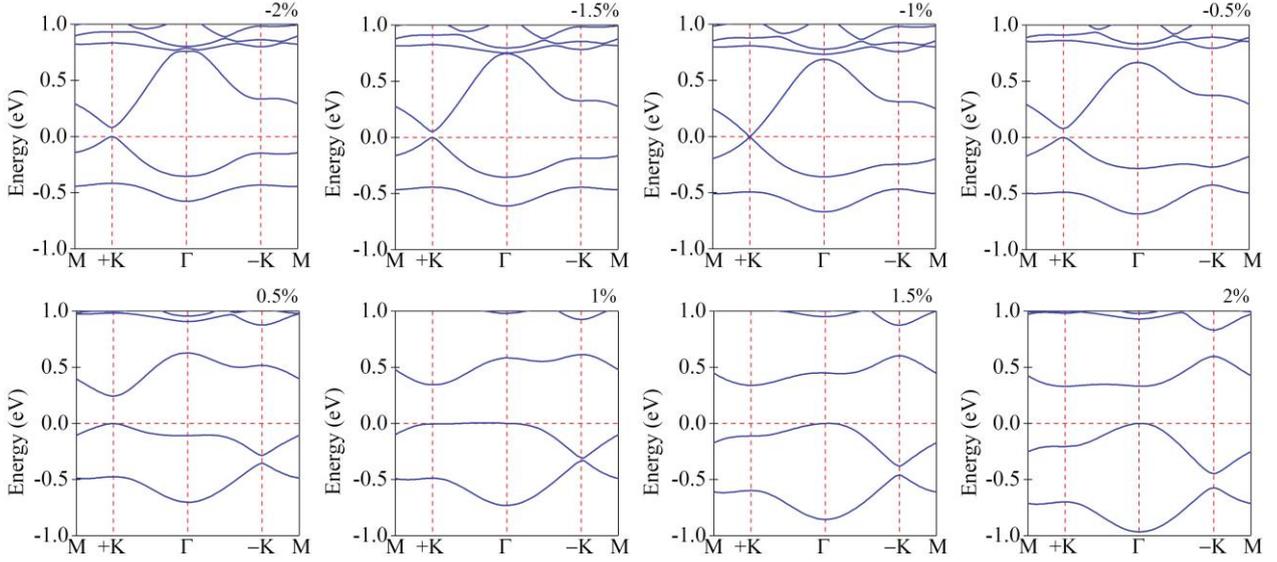

**Fig. 3** Band structures of NbTa-benzene with considering SOC under various strain. The Fermi level is set to 0 eV.

To explore more physics in NbTa-benzene, we investigate the strain effect. The band structures of NbTa-benzene under various strain ranging from -2% to 2% are shown in **Fig. 3**, and the corresponding band gaps at ±K valley are summarized in **Fig. 4(a)**. It can be seen that by varying strain from -2% to 2%, the band gap at –K valley increases from 0.48 to 1.05 eV [**Fig. 4(a)**], which implies that the band order at –K valley is robust in terms of strain. While for +K valley, the band gap decreases from 0.08 to 0 eV with increasing strain from -2% to -1%, and then increases from 0 to 0.53 eV with further increasing strain from -1% to 2%. Such characterized gap-close-and-reopen feature gives rise to the band inversion and thus topological phase transition. The critical point for realizing the band inversion is -1% strain. For NbTa-benzene under -1% strain, the band gap disappears at +K valley and remains semiconducting at –K valley, corresponding to the half-valley-metal state. When compressive strain is smaller than -1%, the band orders at both +K and –K valleys are trivial, and the band topology becomes trivial. In this case, the valley-related multiple Hall phase degenerates into AVH phase solely in NbTa-benzene; see **Fig. 4(a)**. To confirm the valley-contrasting physics is preserved under such band inversion, taking NbTa-benzene under strain of -2% as an example, we calculate its Berry curvature. As shown in **Fig. 4(b)**, there is a small negative value around –K point, while there is a sharp positive peak at +K point due to the tiny band gap. When shifting Fermi level between +K and –K valleys in the conduction (valence) band, the AVH effect still can be observed under an in-plane electric field. The corresponding anomalous Hall conductivity is plotted in **Fig. 4(c)**, from which we can see that when Fermi level is shifted between ±K valleys in the conduction (valence) band, a fully valley-polarized Hall conductivity will be generated (marked by the shaded region). On the other hand, for NbTa-benzene under strain ranging from -1% to 1.5%, the VP-QAH effect is always preserved. However, by increasing strain larger than 1.5%, as shown in **Fig. 3**, the VBM and CBM shifts from +K point to Γ point, deforming the valley features in the valence and conduction bands. As the result, the valley-related multiple Hall phase degenerates into the QAH phase solely; see **Fig. 4(a)**.

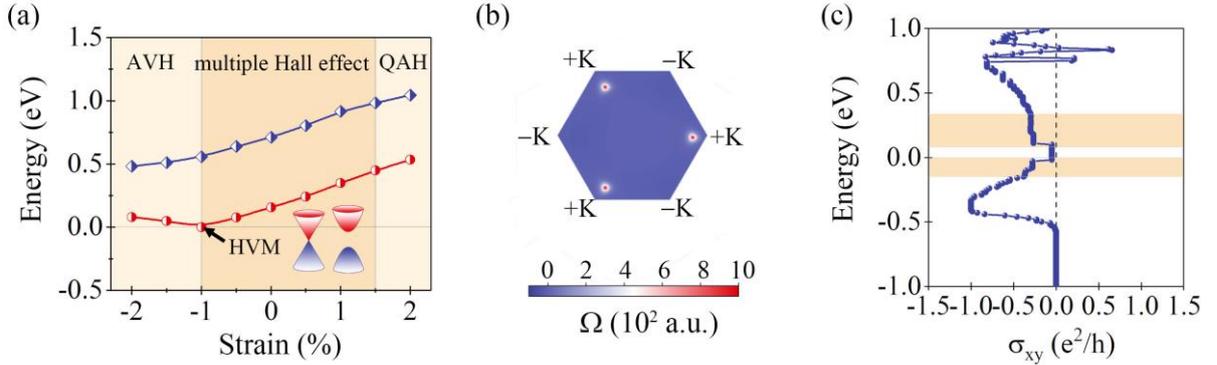

**Fig. 4** (a) Band gap at ±K valley of NbTa-benzene as a function of strain. (b) Contour map of Berry curvature of NbTa-benzene under -2% strain over the entire 2D Brillouin zone. (c) Anomalous Hall conductivity of NbTa-benzene under -2% strain. The shaded regions in (c) highlight the energy levels of the valleys in conduction and valence bands.

## IV. Conclusion

To summarize, we demonstrate the intrinsic valley-related multiple Hall effect in NbTa-benzene through tight-binding model analysis and first-principles calculations. We show that NbTa-benzene is a FM semiconductor with a direct band gap locating at the ±K point. Because of the combined effect of ***T-*** and ***P-***breaking, sizeable valley polarization occurs spontaneously in both the conduction and valence bands, leading to the AVH effect in NbTa-benzene. Moreover, the band orders at +K valley is inversed when including SOC, while the band orders remain trivial at –K valley, which gives rise to the VP-QAH effect in NbTa-benzene. As these two phenomena appear simultaneously, the valley-related multiple Hall effect is successfully achieved. In addition, by applying external, the valley-related multiple Hall effect can be degenerated into either AVH or QAH phases solely, and a general phase diagram is mapped out.

**Acknowledgements**

This work is supported by the National Natural Science Foundation of China (No. 11804190, 12074217), Shandong Provincial Natural Science Foundation (Nos. ZR2019QA011 and ZR2019MEM013), Shandong Provincial Key Research and Development Program (Major Scientific and Technological Innovation Project) (No. 2019JZZY010302), Shandong Provincial Key Research and Development Program (No. 2019RKE27004), Shandong Provincial Science Foundation for Excellent Young Scholars (No. ZR2020YQ04), Qilu Young Scholar Program of Shandong University, and Taishan Scholar Program of Shandong Province.

**Reference**

[1] D. Xiao, G.-B. Liu, W. Feng, X. Xu, and W. Yao, Coupled spin and valley physics in monolayers of $MoS_2$ and other Group-VI dichalcogenides, Phys. Rev. Lett. 108, 196802 (2012).
[2] W. Yao, D. Xiao, and Q. Niu, Valley-dependent optoelectronics from inversion symmetry breaking, Phys. Rev. B 77, 235406 (2008).
[3] D. Xiao, W. Yao, and Q. Niu, Valley-contrasting physics in graphene: magnetic moment and topological transport, Phys. Rev. Lett. 99, 236809 (2007).
[4] J. R. Schaibley, H. Yu, G. Clark, P. Rivera, J. S. Ross, K. L. Seyler, W. Yao, and X. Xu, Valleytronics in 2D materials, Nat. Rev. Mater. 1, 16055 (2016).
[5] K. F. Mak, K. He, J. Shan, and T. F. Heinz, Control of valley polarization in monolayer $MoS_2$ by optical helicity, Nat. Nanotechnol. 7, 494 (2012).
[6] H. Zeng, J. Dai, W. Yao, D. Xiao, and X. Cui, Valley polarization in $MoS_2$ monolayers by optical pumping, Nat. Nanotechnol. 7, 490 (2012).
[7] G. Aivazian, Z. Gong, A. M. Jones, R. Chu, J. Yan, D. G. Mandrus, C. Zhang, D. Cobden, W. Yao, and X. Xu, Magnetic control of valley pseudospin in monolayer $WSe_2$, Nat. Phys. 11, 148 (2015).
[8] Y. C. Cheng, Q. Y. Zhang, and U. Schwingenschlogl, Valley polarization in magnetically doped single-layer transition-metal dichalcogenides, Phys. Rev. B 89, 155429 (2014).
[9] J. Qi, X. Li, Q. Niu, and J. Feng, Giant and tunable valley degeneracy splitting in $MoTe_2$, Phys. Rev. B 92, 121403(R) (2015).
[10] Z. Zhang, X. Ni, H. Huang, L. Hu, and F. Liu, Valley splitting in the van der Waals heterostructure $WSe_2/CrI_3$: The role of atom superposition, Phys. Rev. B 99, 115441 (2019).
[11] W.-Y. Tong, S.-J. Gong, X. Wan, and C.-G. Duan, Concepts of ferrovalley material and anomalous valley Hall effect, Nat. Commun. 7, 13612 (2016).
[12] J. Liu, W.-J. Hou, C. Cheng, H.-X. Fu, J.-T. Sun, and S. Meng, Intrinsic valley polarization of magnetic $VSe_2$ monolayers, J. Phys.: Condens. Matter 29, 255501 (2017).
[13] X. Li, T. Cao, Q. Niu, J. Shi, and J. Feng, Coupling the valley degree of freedom to antiferromagnetic order, Proc. Natl. Acad. Sci. U. S. A. 110, 3738 (2013).
[14] C. Zhang, Y. Nie, S. Sanvito, and A. Du, First-principles prediction of a room-temperature ferromagnetic Janus VSSe monolayer with piezoelectricity, ferroelasticity, and large valley polarization, Nano Lett. 19, 1366 (2019).
[15] C. Luo, X. Peng, J. Qu, and J. Zhong, Valley degree of freedom in ferromagnetic Janus monolayer H-VSSe and the asymmetry-based tuning of the valleytronic properties, Phys. Rev. B 101, 245416 (2020).

[16] R. Peng, Y. Ma, X. Xu, Z. He, B. Huang, and Y. Dai, Intrinsic anomalous valley Hall effect in single-layer $Nb_3I_8$, Phys. Rev. B 102, 035412 (2020).

[17] H. Cheng, J. Zhou, W. Ji, Y. Zhang, and Y. Feng, Two-dimensional intrinsic ferrovalley $GdI_2$ with large valley polarization, Phys. Rev. B 103, 125121 (2021).

[18] P. Zhao, Y. Ma, C. Lei, H. Wang, B. Huang, and Y. Dai, Single-layer $LaBr_2$: Two-dimensional valleytronic semiconductor with spontaneous spin and valley polarizations, Appl. Phys. Lett. 115, 261605 (2019).

[19] Z. He, R. Peng, X. Feng, X. Xu, Y. Dai, B. Huang, and Y. Ma, Two-dimensional valleytronic semiconductor with spontaneous spin and valley polarization in single-layer $Cr_2Se_3$, Phys. Rev. B 104, 075105 (2021).

[20] R. Peng, Z. He, Q. Wu, Y. Dai, B. Huang, and Y. Ma, Spontaneous valley polarization in 2D organometallic lattice, arXiv:2103.16741.

[21] F. D. M. Haldane, Model for a quantum Hall effect without Landau levels: Condensed-matter realization of the "Parity Anomaly", Phys. Rev. Lett. 61, 2015 (1988).

[22] K. He, Y. Wang, and Q. K. Xue, Quantum anomalous Hall effect, Nat. Sci. Rev. 1, 38 (2014).

[23] Z. Liu, G. Zhao, B. Liu, Z. Wang, J. Yang, and F. Liu, Intrinsic quantum anomalous Hall effect with in-plane magnetization: Searching rule and material prediction, Phys. Rev. Lett. 121, 246401 (2018).

[24] C. Niu, N. Mao, X. Hu, B. Huang, and Y. Dai, Quantum anomalous Hall effect and gate-controllable topological phase transition in layered $EuCd_2As_2$, Phys. Rev. B 99, 235119 (2019).

[25] Y. Jin, Z. Chen, B. Xia, Y. Zhao, R. Wang, and H. Xu, Large-gap quantum anomalous Hall phase in hexagonal organometallic frameworks, Phys. Rev. B 98, 245127 (2018).

[26] M. Ezawa, Valley-polarized metals and quantum anomalous Hall effect in silicene, Phys. Rev. Lett. 109, 055502 (2012).

[27] M. Ezawa, Spin valleytronics in silicene: Quantum spin Hall–quantum anomalous Hall insulators and single-valley semimetals, Phys. Rev. B 87, 155415 (2013).

[28] H. Pan, Z. Li, C. Liu, G. Zhu, Z. Qiao, and Y. Yao, Valley-polarized quantum anomalous Hall effect in Silicene, Phys. Rev. Lett. 112, 106802 (2014).

[29] J. Zhang, B. Zhao, and Z. Yang, Abundant topological states in silicene with transition metal adatoms, Phys. Rev. B 88, 165422 (2013).

[30] T. Zhou, J. Zhang, B. Zhao, H. Zhang, and Z. Yang, Quantum spin-quantum anomalous Hall insulators and topological transitions in functionalized Sb(111) monolayers, Nano Lett. 15, 5149 (2015).

[31] T. Zhou, J. Zhang, Y. Xue, B. Zhao, H. Zhang, H. Jiang, and Z. Yang, Quantum spin–quantum anomalous Hall effect with tunable edge states in Sb monolayer-based heterostructures, Phys. Rev. B 94, 235449 (2016).

[32] J. Zhou, Q. Sun, and P. Jena, Valley-polarized quantum anomalous Hall effect in ferrimagnetic honeycomb lattices, Phys. Rev. Lett. 119, 046403 (2017).

[33] T. Zhou, J. Zhang, H. Jiang, I. Žutić, and Z. Yang, Giant spin-valley polarization and multiple Hall effect in functionalized bismuth monolayers, npj Quantum Mater. 3, 39 (2018).

[34] H. Hu, W. Tong, Y. Shen, X. Wan and C. Duan, Concepts of the half-valley-metal and quantum anomalous valley Hall effect, npj Comput. Mater. 6, 129 (2020).

[35] L. Kou, H. Fu, Y. Ma, B. Yan, T. Liao, A. Du, and C. Chen, Two-dimensional ferroelectric topological insulators in functionalized atomically thin bismuth layers, Phys. Rev. B 97, 075429 (2018).

[36] X. Feng, X. Xu, Z. He, R. Peng, Y. Dai, B. Huang, and Y. Ma, Valley-related multiple Hall effect in single-layer $VSi_2P_4$, Phys. Rev. B 104, 075421 (2021).

[37] S. Li, Q. Wang, C. Zhang, P. Guo, and S. A. Yang, Correlation-driven topological and valley states in monolayer $VSi_2P_4$, Phys. Rev. B 104, 085149 (2021).

[38] X. Zhou, R. Zhang, Z. Zhang, W. Feng, Y. Mokrousov, and Y. Yao, Sign-reversible valley-dependent Berry phase effects in 2D valley-half-semiconductors, npj Comput. Mater. 7, 160 (2021).

[39] W. Kohn and L. J. Sham, Self-consistent equations including exchange and correlation effects, Phys. Rev. 140, A1133 (1965).
[40] G. Kresse and J. Furthmüller, Efficient iterative schemes for ab initio total-energy calculations using a plane-wave basis set, Phys. Rev. B 54, 11169 (1996).
[41] J. P. Perdew, K. Burke, and M. Ernzerhof, Generalized gradient approximation made simple, Phys. Rev. Lett. 77, 3865 (1996).
[42] L. Wang, T. Maxisch, and G. Ceder, Oxidation energies of transition metal oxides within the GGA+ U framework, Phys. Rev. B: Condens. Matter 73, 195107 (2006).
[43] H. J. Monkhorst and J. D. Pack, Special points for Brillouin-zone integrations, Phys. Rev. B 13, 5188 (1976).
[44] A. A. Mostofi, J. R. Yates, G. Pizzi, Y.-S. Lee, I. Souza, D. Vanderbilt, and N. Marzari, An updated version of wannier90: A tool for obtaining maximally-localised Wannier functions, Comput. Phys. Commun. 185, 2309 (2014).
[45] Q. Wu, S. Zhang, H. Song, M. Troyer, and A. A. Soluyanov, WannierTools: An open-source software package for novel topological materials, Comput. Phys. Commun. 224, 405 (2018).
[46] J. Zhang, A. Shchyrba, S. Nowakowska, E. Meyer, T. A. Jung, and M. Muntwiler, Probing the spatial and momentum distribution of confined surface states in a metal coordination network, Chem. Commun. 50, 12289 (2014).
[47] P. Liljeroth, I. Swart, S. Paavilainen, J. Repp, and G. Meyer, Single-molecule synthesis and characterization of metal-ligand complexes by low-temperature STM, Nano Lett. 10, 2475 (2010).
[48] D. J. Thouless, M. Kohmoto, M. P. Nightingale, and M. den Nijs, Quantized Hall conductance in a two-dimensional periodic potential, Phys. Rev. Lett. 49, 405 (1982).
[49] Y. G. Yao, L. Kleinman, A. H. MacDonald, J. Sinova, T. Jungwirth, D.-S. Wang, E. Wang, and Q. Niu, First principles calculation of anomalous Hall conductivity in ferromagnetic bcc Fe, Phys. Rev. Lett. 92, 037204 (2004).
[50] T. Cai, S. A. Yang, X. Li, F. Zhang, J. Shi, W. Yao, and Q. Niu, Magnetic control of the valley degree of freedom of massive Dirac fermions with application to transition metal dichalcogenides, Phys. Rev. B 88, 115140 (2013).